\documentclass{article}
\usepackage{graphicx,caption}
\usepackage{latexsym,amsfonts,amsmath,amssymb,mathrsfs} 

\usepackage{subcaption}
\usepackage{float}
\usepackage{hyperref}
\usepackage[english]{cleveref}
\usepackage[nonamebreak,sort&compress]{natbib}
\setcitestyle{numbers,square,comma}
\begin{document}

\date{}
\title{Numerical determination of the  entanglement entropy for a Maxwell field in the cylinder}
\author{Marina Huerta\footnote{e-mail: marina.huerta@cab.cnea.gov.ar} , Leonardo A. Pedraza\footnote{e-mail: leonardo.pedraza@cab.cnea.gov.ar }\\
{\sl Centro At\'omico Bariloche,
8400-S.C. de Bariloche, R\'{\i}o Negro, Argentina}}
\maketitle

\begin{abstract}
We calculate numerically the logarithmic contribution to the entanglement entropy of a cylindrical region in three spatial dimensions for a Maxwell field. Our result does not agree with the analytical predictions concerning any conformal field theory in four dimensions according to which the coefficient is universal and proportional to the type $c$ conformal anomaly.  In cylindrical coordinates the problem decomposes into one dimensional ones along the radial coordinate for each angular momentum. The entanglement entropy of a Maxwell field is equivalent to the one of two identical decoupled scalars with an extra self interaction term. 
\end{abstract}

\section{Introduction}
The coefficient $c_{\log}$ of the logarithmically divergent term in the entanglement entropy (EE) for conformal field theories in four dimensions, has been found to be in general 
a linear combination of the type $a$ and $c$ conformal anomalies  depending on the geometry of the region $V$ \cite{solo}:

\begin{equation}
c_{\log}=\frac{a}{180} \,\chi(\partial V)+\frac{c}{240 \pi}   \int_{\partial V}(k_i^{\mu \nu}k^i_{\nu \mu} -\frac{1}{2} k_i^{\mu \mu}k^i_{\mu \mu})\,.\label{general}
\end{equation}
Here $\chi(\partial V)$ is the Euler number of the surface,  $k^i_{\mu\nu}=-\gamma^\alpha_\mu \gamma^\beta_\nu \partial_\alpha n^i_\beta$ is 
the second fundamental form, $n^\mu_i$ with $i=1,2$ are a pair of unit vectors orthogonal to $\partial V$, and $\gamma_{\mu\nu}=\delta_{\mu\nu}-n^i_\mu n^i_\nu$ is 
the induced metric on the surface.
The sphere and cylinder geometries are particularly interesting to test this result, in the sense that they are sensitive only to one type of anomaly, $a$ or $c$, respectively. From (\ref{general}), it can be seen that only one of the coefficients is non null in each case. This is 
\begin{equation}
c_{log}= \frac{c}{240}\frac{L}{R}\,.
\label{coefcyl}
\end{equation}
for a cylinder of length $L$ and radius $R$ and 
\begin{equation}
c_{log}= \frac{a}{90}\,,
\label{coefsph}
\end{equation}
for spheres.

For spherical sets, this result was later reobtained, and checked numerically and analytically for massless scalars and fermions \cite{num,dowker,cashue,solo2}, using different technics. Moreover,  the identification of the entropy logarithmic coefficient with the Euler type anomaly $a$ was later extended to any dimensions \cite{cashuemye}. 
On the other hand, the result (\ref{coefcyl}) for the cylinder was independently obtained from holographic calculations in four dimensions \cite{Myers}, where it was pointed out that its generalization to higher dimenssion could have a richer structure because of the existence of other conformally invariant terms \cite{Safdi}. Besides, the proportionality with the type $c$ anomaly was succesfully checked numerically for scalars and fermions in \cite{huerta}. 
We remark, that in what concerns gauge fields, specifally the Maxwell field, explicit calculations in the sphere \cite{dowker, coefsph, Maxwell,Maxwell1,Maxwell2,masmaxwell} show conflictive results: using the same standard methods as for scalars and fermions, the $c_{\log}$ does not correspond to the stress tensor anomaly coefficient $a$. Similarly in the context of black holes, already twenty years ago, in the early result \cite{kabat}, Kabat found a non standard negative contribution to the entropy for vector fields, known in the literature as the contact term. On the other hand, in lattice calculations \cite{lattice,masgauge,cashueros}, it was noted the bipartition of the Hilbert space, in the case of gauge fields deserves a special treatment. Some proposals to solve this problem introduce an extended lattice \cite{lattice} or  Hilbert space \cite{masmaxwell}. In these approaches, the Hilbert space admits a bipartition as a tensor product and there are two contributions to the entropy: a classical and a quantum one associated to the boundary and bulk degrees of freedom respectively. Later, the problem of the bipartition of the Hilbert space, was studied from an algebraic point of view \cite{cashueros}. In \cite{cashueros}, this problem was related to a more general one: the identification of local algebras with regions due to the presence of a center, a set of operators which conmute with all the rest and live on the boundary region. The ambiguity in the assignation of local algebras to regions is not a particular issue of gauge fields but it concerns any general case. Regarding the consequences of this ambiguity on different information measures, it was pointed out, that even if these ambiguities in the election of the local density matrix are directly inherited by the EE by an extra classical contribution, the mutual information (MI) is not sensitive to them. Thus, the fine tunning in the election of the center (the electric center choice is equivalent to the extended lattice or Hilbert space approaches) can solve the mismatch in the EE \cite{Maxwell,masmaxwell} but not in the MI. Another example of this kind are the topological theories \cite{kitaev, wen}, where as discussed in \cite{gaugecashue} also present conflictive results in the identification of the topological coefficient and the constant term in the EE expansion in two spatial dimensions that can be solved at the EE level but not in the MI. 

In this context, we study here the case of the EE for the Maxwell field in the cylinder.

We list below, the expected values of $c_{log}$ for Maxwell, scalar and fermion free fields
\begin{align}\label{cyl}
c^{M}_{\log}&=\frac{12L}{240R}\,,\\ 
c^{s}_{\log}&=\frac{L}{240R} \,,\\
c^{f}_{\log}&=\frac{6L}{240R}\,.
\end{align}

As we said before, this result was succefully checked for scalar and fermion fields in \cite{huerta}. The method used in \cite{huerta} consists first, in dimensionally reducing the three dimensional cylinder problem to the one of an infinite set of massive fields living in a two dimensional spatial disk. In the disk, the cylinder logarithmic coefficient corresponds to the one of $(mR)^{-1}$ expanding the entanglement entropy in powers of $mR$. Here, we follow the same steps as in \cite{huerta}: we first dimensionally reduce the problem considering the cylinder length $L$ is much larger than the radius $R$. Then, the problem in the disk, by rotational symmetry,  ends up in a one dimensional problem in the radial coordinate. We find the original problem corresponds to two identical decoupled scalar fields with a local quadratic self interaction term. 
The EE logarithmic coefficient for this model is finally calculated numerically in a radial lattice, using the method described in \cite{review, sred, peschel} where the entropy is written in terms of vacuum two points correlators. We find $c^M_{\log}$ does not agree with the analytical prediction (\ref{cyl}).

\section{Scalar and Maxwell field in a cylinder}
We are interested in calculating the logarithmic contribution to the EE of the Maxwell field in a cylindrical region of length $L$ and radius $R$ in three spatial dimensions. 
For free fields, it is possible to dimensionally reduce the problem when $L\gg R$ \cite{review}, 
keeping the dominant extensive term invariant. We will profit of this, to map the problem from the cylinder to the disk. 
This section is organized as follows: We start studying the scalar field case in a cylindrical region where the dimensional reduction method is briefly introduced. Then, we generalize the scalar case to vector fields.

\subsection{Scalar field}
The EE for a massless scalar field in a cylinder was studied in detail in \cite{huerta}. For completeness, we repeat here some of the formulae, since they are going to be useful in generalizing the problem to the vector field case and give an interpretation to the Maxwell result.

 The Hamiltonian  of a massless scalar field is 
 \begin{equation}
     H=\frac{1}{2}\int dV \left(\pi^{2}+ \nabla\phi^{2}\right).
 \end{equation}
Since we are considering a cylindrical region, the field $\phi(\textbf{x})$ can be conveniently express as
   \begin{equation}
     \phi(\rho,\theta,z)=\sum_{n} \frac{1}{\sqrt{ L}} e^{i\frac{2\pi n}{L} z} \phi_{n}(\rho,\theta),\label{field}
 \end{equation}
where $(\rho,\theta,z)$ are cylindrical coordinates. In (\ref{field}), we have imposed periodical boundary conditions $z\equiv z+L$ by compactifying the direction $z$ along the cylinder axis. The fields $\phi_{n}(\rho,\theta)$ 
 defined in the disk, are associated to the each $n$ axial mode. 
 As explained in \cite{review}, the dimensional reduction gives rise to an emerging  mass for the fields in the disk given by $m=k_n=2\pi n/L$ which can be read directly from the Hamiltonian
 \begin{equation}
    H=\sum_{n} \int d\theta\, d\rho  \,\rho \left( \pi_{n}^{2}
    +(\nabla_{D}\phi_{n})^{2}+k_n^{2}\phi_{n}^{2} \right).
\end{equation}
Here $\nabla_{D}$ is the bidimensional gradient in the disk and  $\pi_{n}(\rho, \theta)=\partial_{0}\phi_{n}(\rho, \theta)$ is the conjugated canonical momentum of the field. 

Under the condition $L\gg R$, the cylinder EE  is related to the disk EE by an integral over the  fields mass
\begin{equation}
    S(V)=\frac{L}{\pi}\int_{0}^{\infty}dm\,S\left(D,m\right).
\label{eq:S_cylinder}
\end{equation}

From (\ref{eq:S_cylinder}), it is easy to see that there is a one to one identification between coefficients of the cylinder and disk EE expansions. In particular, for the logarithmic coefficient, we have
\begin{equation}
    c_{log}=-c_{(-1)}\frac{L}{\pi R}\,,\label{relacioncoef}
\end{equation}
where $c_{(-1)}$ is read off from the expansion of $S\left(D,m\right)$  in powers of $mR$ 
\begin{equation}
    S\left(D,m\right)=c_{1}mR+c_{0}+c_{(-1)}\frac{1}{mR}+...\,.
    \label{expansion_de_s_en_mr}
\end{equation}

Moreover,  the resulting problem in the disk can still be dimensionally reduced once again due to the rotational symmetry. Thus, taking into account the rotational symmetry, the field (\ref{field}) admits a decomposition in angular modes    
 \begin{equation}\label{fourier_decomposition}
     \phi(\rho,\theta,z)=\sum_{n=-\infty}^{\infty}\sum_{l=-\infty}^{\infty} \frac{1}{\sqrt{2\pi L}} e^{i l \theta} e^{i\frac{2\pi n}{L} z} \phi_{l\,n}(\rho).
 \end{equation}

The fields $\phi_{l\,n}$ satisfy
 \begin{equation}\label{realidad_l}
     \phi_{l\,n}=\phi_{l\,n}^{\dagger}=\phi_{-l\,n}.
 \end{equation}
for each $n$ mode.
Then, introducing a set of functions
 \begin{equation}\label{base_fourier}
     f_{l\,n}(\theta,z)=\frac{1}{\sqrt{2\pi L}} e^{i l \theta} e^{i\frac{2\pi n}{L} z},
 \end{equation}
which are orthonormal by construction 
 \begin{equation}\label{ortonormalidad}
     \int_{-\pi}^{\pi} d\theta \int_{-L/2}^{L/2} dz f_{l\,n}(\theta,z)f_{l'\,n'}^{*}(\theta,z)=\delta_{l\,l'}\delta_{n\,n'},
 \end{equation}
the field and its conjugated momentum take the form

 \begin{equation}\label{redefinicion}
     \phi(\rho,\theta,z)=\sum_{n,l} f_{l\,n}(\theta,z) \phi_{l\,n}(\rho)\,,
 \end{equation}

 \begin{equation}\label{momentoconjugado}
     \pi(\rho,\theta,z)=\partial_{0} \phi(\rho,\theta,z)=\sum_{n,l} f_{l\,n}(\theta,z) \pi_{l\,n}(\rho)\,,
 \end{equation}
 with $\pi_{l\,n}=\partial_{0}\phi_{l\,n}$.
 The commutation relation between fields and momenta reads in cylindrical coordinates
\begin{equation}
    [\phi(\rho,\theta,z),\pi(\rho',\theta',z')]=\frac{i}{\rho} \delta(\rho-\rho')\delta(\theta-\theta')\delta(z-z') \,.
\end{equation}

In order to restore canonical commutation relation, we rescale the field and momentum modes as follows
\begin{equation}\label{camposreescaliados}
   \tilde{\phi}_{l\,n}=\rho^{1/2}\phi_{l\,n}\,\,;\, \, \tilde{\pi}_{l\,n}=\rho^{1/2}\pi_{l\,n}, 
\end{equation}
The new set $(\tilde{\phi}_{l\,n},\tilde{\pi}_{l\,n})$ satisfies 
\begin{equation}
    [\tilde{\phi}_{l\,n}(\rho),\tilde{\pi}_{l'\,n'}(\rho')]=i \delta_{l\,l'} \delta_{n\,n'} \delta(\rho-\rho').
\end{equation}
 
Finally, the Hamiltonian is reduced to the sum
 \begin{equation}
     H=\sum_{n=-\infty}^{\infty} H_{n}=\sum_{n=-\infty}^{\infty}\sum_{l=-\infty}^{\infty} H_{l\,n},
 \end{equation}
 where
 \begin{equation}\label{eq:H_radial}
     H_{l\,n}=\frac{1}{2}\int d\rho \left\{ 
     \left( \tilde{\pi}_{l\,n}\right)^2+
     \rho \left[\frac{\partial}{\partial\rho} \left( \rho^{-1/2}\tilde{\phi}_{l\,n}\right) \right]^2+
     \left[\frac{l^2}{\rho^2}+k_n^2\right]\left(\tilde{\phi}_{l\,n}\right)^2\right\}.
 \end{equation}
For each mode $n$, the uncoupled fields $\tilde{\phi}_{l\,n}(\rho)$ will contribute extensively to the EE
  \begin{equation}
  S=S_0+2\sum_{l=1}^{\infty}S_l \,. \label{sumentropy}
  \end{equation}
There is a factor $2$ in the sum over $l$ in \cref{sumentropy}. This is because  modes $l$ and $-l$ give identical contributions. 
The original problem in $(3+1)$ dimensions is finally reduced to another one in $(1+1)$ dimensions. In the Section \ref{hamiltoniano_campo_vectorial} we generalize this result to the vector field case. 

\subsection{Maxwell field}
 \label{hamiltoniano_campo_vectorial}
 
The EE of a Maxwell field in the cylinder can be calculated following the same steps as in the previous case.   
Following the same strategy as in \cite{coefsph} for a spherical region, we start with the Maxwell field Hamiltonian written in terms of the physical electric and magnetic fields
\begin{equation}
    H=\frac{1}{2} \int dV \left(\textbf{E}^2+\textbf{B}^2\right)\,.
\end{equation}

The vectors $\textbf{E}$ and $\textbf{B}$ can be written in cylindrical coordinates in terms of the base functions $f_{l\,n}(\theta,z)$ (\ref{base_fourier})
\begin{equation}\label{campoenpolaresE}
    \textbf{E}=\left\{E^{\rho}_{l\,n}(\rho) \hat{\rho} + E^{\theta}_{l\,n}(\rho) \hat{\theta} + E^{z}_{l\,n}(\rho) \hat{z} \right\} f_{l\,n}(\theta,z)\,,
\end{equation}
\begin{equation}\label{campoenpolaresB}
    \textbf{B}=\left\{B^{\rho}_{l\,n}(\rho) \hat{\rho} + B^{\theta}_{l\,n}(\rho) \hat{\theta} + B^{z}_{l\,n}(\rho) \hat{z} \right\} f_{l\,n}(\theta,z)\,.
\end{equation}
In the vacuum, the fields satisfy the constraints 
\begin{equation}\label{constrain1}
    \nabla \cdot \textbf{E} =\nabla \cdot \textbf{B}=0
\end{equation}
which, for example, give for the electric field
\begin{equation}
    \frac{1}{\rho} \frac{\partial}{\partial\rho} 
    \left(\rho E^{\rho}_{l\,n} \right)+i\frac{l}{\rho}E^{\theta}_{l\,n}+i k_nE^{z}_{l\,n}=0\,.
\end{equation}

\begin{figure}[t]
	\includegraphics[width=0.25\textwidth]{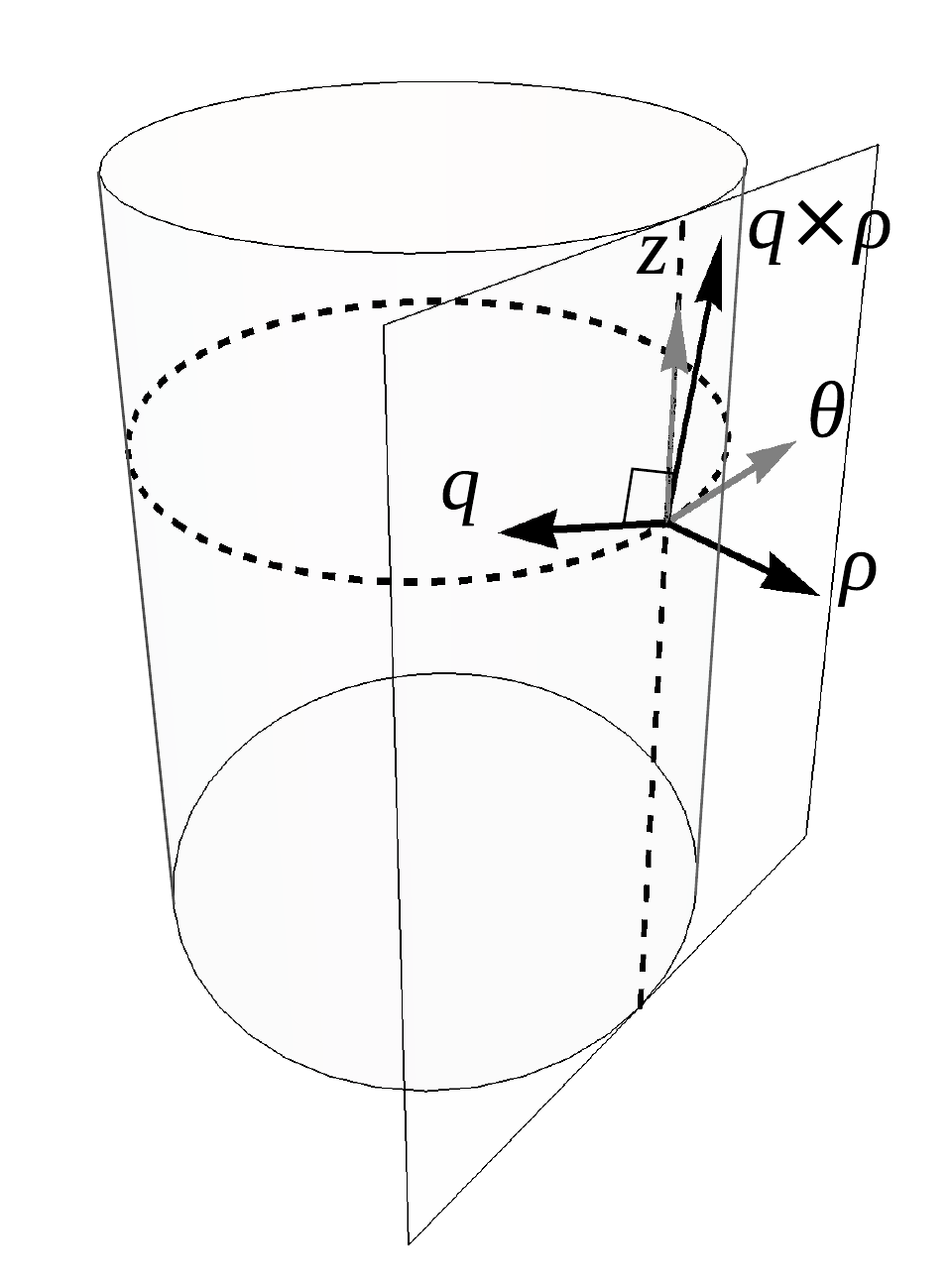}
	\centering
	%\captionsetup{width=.5\linewidth}
	\caption{Relation between bases $(\hat{\rho},\hat{\theta},\hat{z})$ and $(\hat{\rho},\hat{q},\hat{q}\times\hat{\rho})$ used to write the Maxwell field Hamiltonian (\ref{hamiltoniano_maxwell_1}).	\label{fig:cilindro_coords}
	}  
\end{figure}

The above constraint equations take a particular simple form by introducing a new coordinates base.  
We consider the vector $\textbf{q}_n$ defined as
\begin{equation}
\textbf{q}_n=(0,l/\rho,k_n) \,.
\end{equation}
with $k_n=2\pi n/L$.
In this base, the electric and magnetic vectors admit a decomposition analogue to (\ref{campoenpolaresE}) and (\ref{campoenpolaresB}). For the electric field this is
\begin{equation}\label{paralellperpdecomposition}
    \textbf{E}=\left\{E^{\rho}_{l\,n}(\rho) \hat{\rho} + E_{l\,n}^{\parallel}(\rho) \hat{q}_n + E_{l\,n}^{\perp}(\rho) (\hat{q}_n\times\hat{\rho}) \right\} f_{l\,n}(\theta,z),
\end{equation}
with
\begin{equation}
        E_{l\,n}^{\parallel}=\frac{1}{|\textbf{q}_n|}\left(\frac{l}{\rho}E^{\theta}_{l\,n} +k_n E^{z}_{l\,n} \right)\,,
\end{equation}
and 
\begin{equation}
        E_{l\,n}^{\perp}=\frac{1}{|\textbf{q}_n|}\left(k_n E^{\theta}_{l\,n} -\frac{l}{\rho} E^{z}_{l\,n} \right)\,.
\end{equation}
Here,  $E_{l\,n}^{\parallel}$ is parallel to $\textbf{q}_n$ and  $E_{l\,n}^{\perp}$ orthogonal to $\textbf{q}_n$ and $\hat{\rho}$. The same applies to $\textbf{B}$.
The previous constraint now gives the following relation between the parallel and radial field components
\begin{align}\label{constraint}
	&E_{ln}^{\parallel}=\frac{i}{|\textbf{q}_n|} \frac{1}{\rho} \frac{\partial}{\partial\rho} 
	\left(\rho E^{\rho}_{ln} \right) \,,
	& &B_{ln}^{\parallel}=\frac{i}{|\textbf{q}_n|} \frac{1}{\rho} \frac{\partial}{\partial\rho} 
	\left(\rho B^{\rho}_{ln} \right)\,.
\end{align}

In this base and using \cref{ortonormalidad,constraint}, the Hamiltonian reads
\begin{equation}
     H=\sum_{n=-\infty}^{\infty} H_{n}=\sum_{n=-\infty}^{\infty} \sum_{l=-\infty}^{\infty} H_{l\,n},
 \end{equation}
 where
\begin{equation} \label{hamiltoniano_maxwell_1}
        H_{l\,n}= \frac{1}{2} \int d\rho \rho
     \left\{  \left(E^{\rho}_{l\,n} \right)^2 + \frac{1}{\textbf{q}_n^2 \rho} \left[\frac{\partial}{\partial\rho} 
    \left(\rho E^{\rho}_{l\,n} \right) \right]^2 + \left( B^{\perp}_{l\,n} \right)^2 + (E_{l\,n} \leftrightarrow B_{l\,n}) \right\}.
\end{equation}

As in the scalar case, we have reduced the original problem from $(3+1)$ dimensions to another one in $(1+1)$ dimensions. 
We note the constraint (\ref{constrain1}) has removed two degrees of freedom. This explains why in (\ref{hamiltoniano_maxwell_1})  just four of the original six components of the electromagnetic field contribute. 

As in \cite{coefsph}, there is a correspondence between the fields in (\ref{hamiltoniano_maxwell_1}) and the field $\tilde{\phi}$ and its canonical conjugated momentum $\tilde{\pi}$ in (\ref{eq:H_radial}). To proceed with this identification , we start computing the commutation relations satisfied by the electric and magnetic fields.

 In Cartesian coordinates, the commutators between electric and magnetic fields are 
\begin{equation}\label{comnutadorescartesinos}
    \left[E^{i}(\textbf{x}),B^{j}(\textbf{x}')\right]=-i\epsilon^{i j k}\partial_{k}\delta^{3}(\textbf{x}-\textbf{x}')
\end{equation}
which, after the coordinates transformation, can be rewritten as
\begin{align}\label{cnmutadorespolaresfinal}
	\left[E^{\rho}_{l\,n}(\rho),{B^\dagger}^{\theta}_{l'\,n'}(\rho')\right]=  &  \frac{k_n}{\rho}
	\delta_{l\,l'}\delta_{n\,n'}\delta(\rho-\rho'),  \nonumber \\
	\left[E^{\rho}_{l\,n}(\rho),{B^\dagger}^{z}_{l'\,n'}(\rho')\right] = & - \frac{l}{\rho^2} \delta_{l\,l'}\delta_{n\,n'}
	\delta(\rho-\rho'), \\
	\left[E^{\theta}_{l\,n}(\rho),{B^\dagger}^{z}_{l'\,n'}(\rho')\right]= & -\frac{i}{\rho} 
	\delta_{l\,l'}\delta_{n\,n'}\frac{\partial}{\partial \rho}\delta(\rho-\rho')\,. \nonumber 
\end{align}

Due to the constraint equation (\ref{constrain1}), there is only one independent commutator  $\left[E^{\rho}_{l\,n}(\rho),{B^\dagger}^{\theta}_{l'\,n'}(\rho')\right]$. Thus, we can identify two pairs $(\phi^{1,2},\pi^{1,2})$
\begin{align}
	\phi^{(1)}_{l\,n}&=\frac{1}{|\textbf{q}_n|}  \rho^{1/2} E^{\rho}_{l\,n}\,, &
	\pi^{(1)}_{l\,n}&=-i \rho^{1/2} B^{\perp}_{l\,n}\,,\\
	\phi^{(2)}_{l\,n}&=\frac{1}{|\textbf{q}_n|}  \rho^{1/2} B^{\rho}_{l\,n} \,,&
	\pi^{(2)}_{l\,n}&=-i \rho^{1/2} E^{\perp}_{l\,n}\,.
\end{align}
Using \cref{cnmutadorespolaresfinal}, we check the fields and conjugated momenta defined above satisfy canonical commutation relations
\begin{align}\label{cnmutadoresescalares2}
	\left[\phi^{(1)}_{l\,n}(\rho),{\pi^{(1)}_{l'\,n'}}^\dagger(\rho')\right]&=  i \delta_{l\,l'}\delta_{n\,n'}\delta(\rho-\rho'); & 
	\left[\phi^{(2)}_{l\,n}(\rho),{\pi^{(2)}_{l'\,n'}}^\dagger(\rho')\right]&= i \delta_{l\,l'}\delta_{n\,n'}\delta(\rho-\rho').
\end{align}

The Hamiltonian (\ref{hamiltoniano_maxwell_1}) in terms of the new fields reads
\begin{equation}\label{eq:hamiltoniano_maxwell_2}
	\begin{split}
	H_{l\,n}=\frac{1}{2} \int d\rho \bigg\{  \left(\pi^{(1)}_{l\,n}\right)^2+\frac{1}{\textbf{q}_n^2 \rho} \left[\frac{\partial}{\partial \rho}\left( |\textbf{q}_n| \rho^{1/2} \phi^{(1)}_{l\,n}\right)\right]^2+
	\textbf{q}_n^2 \left(\phi^{(1)}_{l\,n} \right)^2\\  
	+ \left(\pi^{(2)}_{l\,n}\right)^2+\frac{1}{\textbf{q}_n^2 \rho} \left[\frac{\partial}{\partial \rho}\left( |\textbf{q}_n| \rho^{1/2} \phi^{(2)}_{l\,n}\right)\right]^2+
	\textbf{q}_n^2 \left(\phi^{(2)}_{l\,n} \right)^2\bigg\}.
	\end{split}
\end{equation}
We note that in (\ref{eq:hamiltoniano_maxwell_2}), there are two identical copies labeled by the indices $1,2$. 
In order to compare the Hamiltonian for each copy with (\ref{eq:H_radial}) found in the previous section for a scalar field, we rewrite the second term  $\frac{1}{\textbf{q}_n^2 \rho} \left[\frac{\partial}{\partial \rho}\left( 	|\textbf{q}_n| \rho^{1/2} \phi_{l\,n}\right)\right]^2$ as
\begin{equation}
	\begin{split}
	\frac{1}{\textbf{q}_n^2 \rho} \left[\frac{\partial}{\partial \rho}\left( 	|\textbf{q}_n| \rho^{1/2} \phi_{l\,n}\right)\right]^2=
	\rho \left[\frac{\partial}{\partial\rho}\left(\rho^{-1/2}\phi_{l\,n}\right) \right]^2 
	+\frac{\partial}{\partial\rho} \left[ \frac{k_n^2}{|\textbf{q}_n|^2 \rho} \left(\phi_{l\,n} \right)^2 \right]\\
	+\frac{\rho^2 k_n^4 -2 l^2 k_n^2}{(\rho^2 k_n^2 + l^2)^2} \left(\phi_{l\,n} \right)^2
	\end{split} \,.\label{derivativeterm}
\end{equation} 
From (\ref{derivativeterm}), we conclude the Maxwell field in the cylinder except for a total derivative term, is equivalent to two decoupled scalar fields $\phi_1$ and $\phi_2$ with a local self interaction term 
\begin{equation}\label{autointeraccion}
    V(\phi^p)=\sum_{l,n} \frac{\rho^2 k_n^4 -2 l^2 k_n^2}{(\rho^2 k_n^2 + l^2)^2} \left(\phi^p_{l\,n} \right)^2\,,\,\,\,\,\,\,p=1,2.
\end{equation}

The EE as before will be given for each mode $n$ by the sum over $l$ of the independent contributions $S_l$ (\ref{sumentropy}). 
In Section \ref{secnumerics} we calculate $S$ numerically in a radial lattice.

\section{Numerical calculation: The system in the lattice}
\label{secnumerics}
The EE for scalars can be written as  \cite{review}
\begin{equation}\label{eq:entropy}
	S=\textrm{Tr}\left((C+1/2)\textrm{ln}(C+1/2)-(C-1/2)\textrm{ln}(C-1/2)\right)
\end{equation} 
where $C=\sqrt{X P}$ and $X$ and $P$  are real, symmetric and positive matrices corresponding to the vacuum two points correlators 
\begin{align}
	&\langle\phi_{i}\phi_{j}\rangle=X_{ij}, & &\langle\pi_{i}\pi_{j}\rangle=P_{ij}.
\label{2}
\end{align}
Here the indices correspond to lattice positions restricted to the considered region.  On the other hand, for a general Hamiltonian of the form
\begin{equation}
H=\frac{1}{2}\sum_{i}\pi_{i}^{2}+\frac{1}{2}\sum_{i,j}\phi_{i}K_{ij}\phi_{j},
\end{equation}
the vacuum correlators are related to $K$ by
\begin{align}\label{eq:correladors_XandP}
&X_{ij}=\frac{1}{2}\left(K^{-1/2}\right)_{ij},
&P_{ij}=\frac{1}{2}\left(K^{1/2}\right)_{ij}.
\end{align} 
In our case, the matrix  $K_{Maxwell}$ has to be read from the Hamiltonian (\ref{eq:hamiltoniano_maxwell_2})  in its discrete version (for a given mass $m$ and dropping the $l$ index for simplicity)
\begin{equation}\label{eq:discrete_Maxwell_Hamiltonian}
H=\left\{\frac{1}{2}\sum_{i}\pi_{i}^{2}+\frac{1}{2}\sum_{i}\left\{\left(i+1/2\right)\left(\frac{\phi_{i+1}}{\sqrt{i+1}}-\frac{\phi_{i}}{\sqrt{i}}\right)^{2}+\frac{l^{2}}{i^{2}}\phi_{i}^{2}+m^{2}\phi_{i}^{2}+\left[\frac{i^2 m^4-2 m^2 l^2}{(l^2 +i^2 m^2)^2}\right]\phi_{i}^{2}\right\}\right\},
\end{equation}
which gives

\begin{align}\label{eq:K^ij(2)}
    &K_{m,l}^{1,1}=\frac{3}{2}+l^{2}+m^{2}
    +\frac{m^4-2 m^2 l^2}{(l^2 + m^2)^2}, \nonumber \\ 
    &K_{m,l}^{i,i}=2+\frac{l^{2}}{i^{2}}+m^{2} +\frac{i^2 m^4-2 m^2 l^2}{(l^2 +i^2 m^2)^2},  \\     &K_{m,l}^{i+1,i}=K_{m,l}^{i,i+1}=-\frac{i+1/2}{\sqrt{i\left(i+1\right)}}. \nonumber
\end{align}

This is the same matrix found in \cite{huerta} for scalars, except for extra diagonal terms which come from the self interaction term. We note that the Maxwell Hamiltonian corresponds to two identical copies of (\ref{eq:discrete_Maxwell_Hamiltonian}).

As  it is mentioned in the previous section (and extensively discussed in \cite{huerta} and \cite{review}), due to the dimensional reduction procedure, the  cylinder EE is directly related to the disk EE giving a correspondence between coefficients of the respective expansions (\ref{relacioncoef}). 
The problem in the disk, is in turn reduced to a one dimensional one in the radial coordinate because of the rotational symmetry. 
For the calculation, we have used a one dimensional (radial) lattice of $500$ points size. We have chosen to work with a fixed lattice size after checking there were no significative IR dependence in our results. We have considered disc sizes with radii within the range $100 \leq R\leq 350$ and  masses $1/15 \leq m \leq1/5$, in lattice units. 
We compute the matrix $K$ for different pairs $(m, l)$. This is a $500\times 500$ matrix. For a given mass, this has to be done for different angular momentum values. We have considered $0\leq l\leq l_{max}$ with $l_{max}=3000$.  
Finally,  from (\ref{eq:correladors_XandP}) we obtain $C=C_{l}(R,m)$ and in turn $S_{l}(R,m)$ through (\ref{eq:entropy}).

The total entropy  for a given mass and disk size is obtained summing over the angular momentum contributions $S_l$
\begin{align}
	S(R,m)=S_{0}+\sum_{l=1}^{l_{max}}2S_{l}+\mathcal{O}\left(S_{l_{max}}\right).\label{sumentrolattice}
\end{align}
We calculate $S_l$ exactly up to a maximum angular momentum $l_{max}=3000$. In (\ref{sumentrolattice}), $\mathcal{O}\left(S_{l_{max}}\right)$ is a correction that takes into account large $l$ contributions for $l>l_{max}$. 
As in \cite{huerta}, we estimate the large angular momentum correction using the following fit
\begin{equation}
	S_{l}=\frac{1}{l^{2}}+\frac{\ln(l)}{l^{2}}+\frac{1}{l^{4}}+\frac{\ln(l)}{l^{4}}+\frac{1}{l^{6}}+\frac{\ln(l)}{l^{6}}+\frac{1}{l^{8}}+\frac{\ln(l)}{l^{8}}.
\end{equation}
\begin{figure}[t]
	\includegraphics[width=0.75\textwidth]{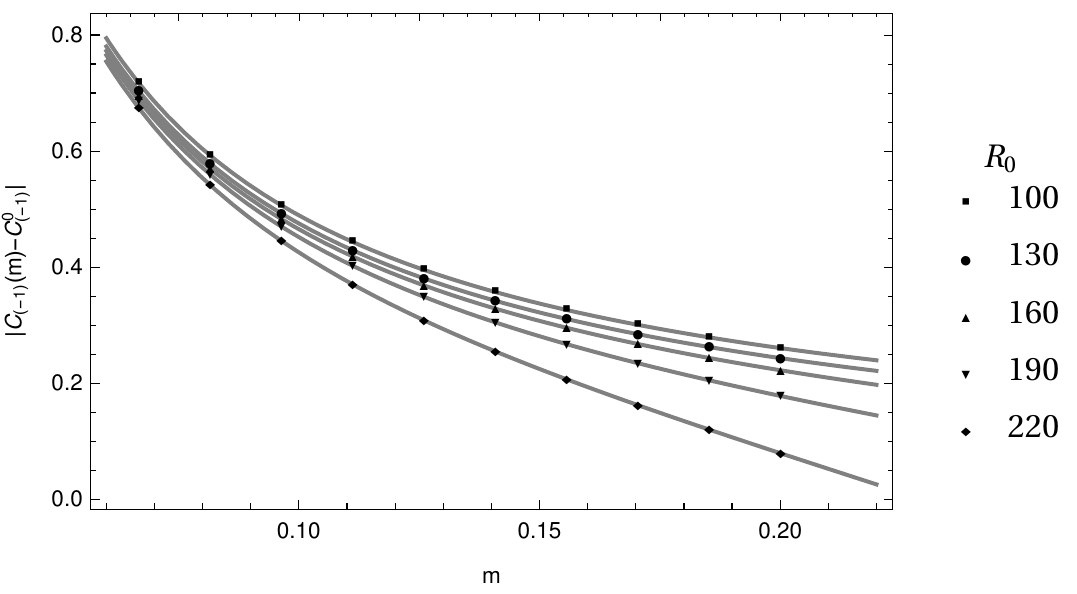}
	\centering
	%\captionsetup{width=.75\linewidth}
	\caption{\label{fig:M_C_-1}
		The points correspond to $\vert c^{(R_0)}_{(-1)}(m)-c^{(0,R_0)}_{(-1)}\vert$ where $c^{(R_0)}_{(-1)}(m)$ is the coefficient of the term proportional to $\frac{1}{R}$ in the EE expansion (\ref{fitRR0}) (the constant term has been substracted for ease of visualization).These coefficients were obtained fitting the entropy within different intervals   $R_0<R<\frac{3}{2}R_0$ with $R_0=100,130,160,190,220$ from top to bottom, as explained in the text. The solid lines correspond to the fit (\ref{fitmR0}) where it is apparent the $\frac{1}{m}$ behavior of the curve. The coefficient of the term proportional to $\frac{1}{m}$ is related to the logarithmic coefficient by (\ref{result_maxwell}).}  
\end{figure} 
Finally, we extract the $c_{(-1)}$ coefficient from $S(m_i,R_j)$ in three steps. 

First, keeping the mass $m_j=m$ fixed, we fit the pairs $(R_j, S(R_j,m))$ as
\begin{equation}
	S(R,m)=c_{(-1)}(m)\frac{1}{R}+c_{0}(m)+c_{1}(m)R\,.
	\label{fitR}
\end{equation}
Then, we extract from (\ref{fitR}) the coefficient $c_{(-1)}(m_j)$ of the term proportional to $\frac{1}{R}$  and fit the pairs $(m_j,c_{(-1)}(m_j))$  with a power expansion of the form 
\begin{equation}
	c_{(-1)}(m)=c_{(-1)}\frac{1}{m}+c_{(-1)}^{0}+c_{(-1)}^{1}m+c_{(-1)}^{2}m^2+c_{(-1)}^{3}m^3\,.
	\label{fitm}
\end{equation}
For this fit we have used $10$ equispaced masses from $1/15$ to $1/5$ in lattice units. Note that in (\ref{fitm}), only the first term is universal and the others contain positive powers of the cut-off which is $\epsilon=1$ here.
To obtain maximal presicion, we proceed as follows.
The fits above, are repetead within different intervals: $R_{0}\leq R_j\leq R_{f}=3/2\,R_{0}$ and $100\leq R_{0}\leq 230$. We labeled each \textit{interval} with the smaller radios  value $R_0$. Thus, we have one coefficient $c^{R_0}_{(-1)}(m)$ for each window $R_{0}$. For clarification purposes, we repeat here the corresponding power expansion
\begin{equation}
	S^{R_0}(R,m)=c^{R_0}_{(-1)}(m)\frac{1}{R}+c^{R_0}_{0}(m)+c^{R_0}_{1}(m)R\,,
	\label{fitRR0}
\end{equation}
\begin{equation}
	c^{R_0}_{(-1)}(m)=c^{R_0}_{(-1)}\frac{1}{m}+c_{(-1)}^{(0,R_0)}+c_{(-1)}^{(1,R_0)}m+c_{(-1)}^{(2,R_0)}m^2+c_{(-1)}^{(3,R_0)}m^3\,.
	\label{fitmR0}
\end{equation}
In \cref{fig:M_C_-1}, we plot $\vert c^{R_0}_{(-1)}(m)-c_{(-1)}^{(0,R_0)}\vert$ for different intervals  with $R_0=100,130,160,190, 220$ (the constant term has been substracted for ease of visualization). For example, for $m=\frac{1}{9}$ and $R_0=100$, the expansion
(\ref{fitRR0}) is $-4.56671\frac{1}{R}+ 0.2912071+0.4092523R$, given  $c^{R_0}_{(-1)}(\frac{1}{9})=-4.56671$. In the same figure, the solid lines correspond to the fit (\ref{fitmR0}). For example, for $R_0=100$, we obtain, $ - 0.0466873\frac{1}{m} - 4.12286 - 0.364247\, m + 1.64364\, m^2 - 2.52775\, m^3$, resulting $c^{R_0=100}_{(-1)}= -0.0466873$.

\begin{figure}[t]	
	\includegraphics[width=0.75\textwidth]{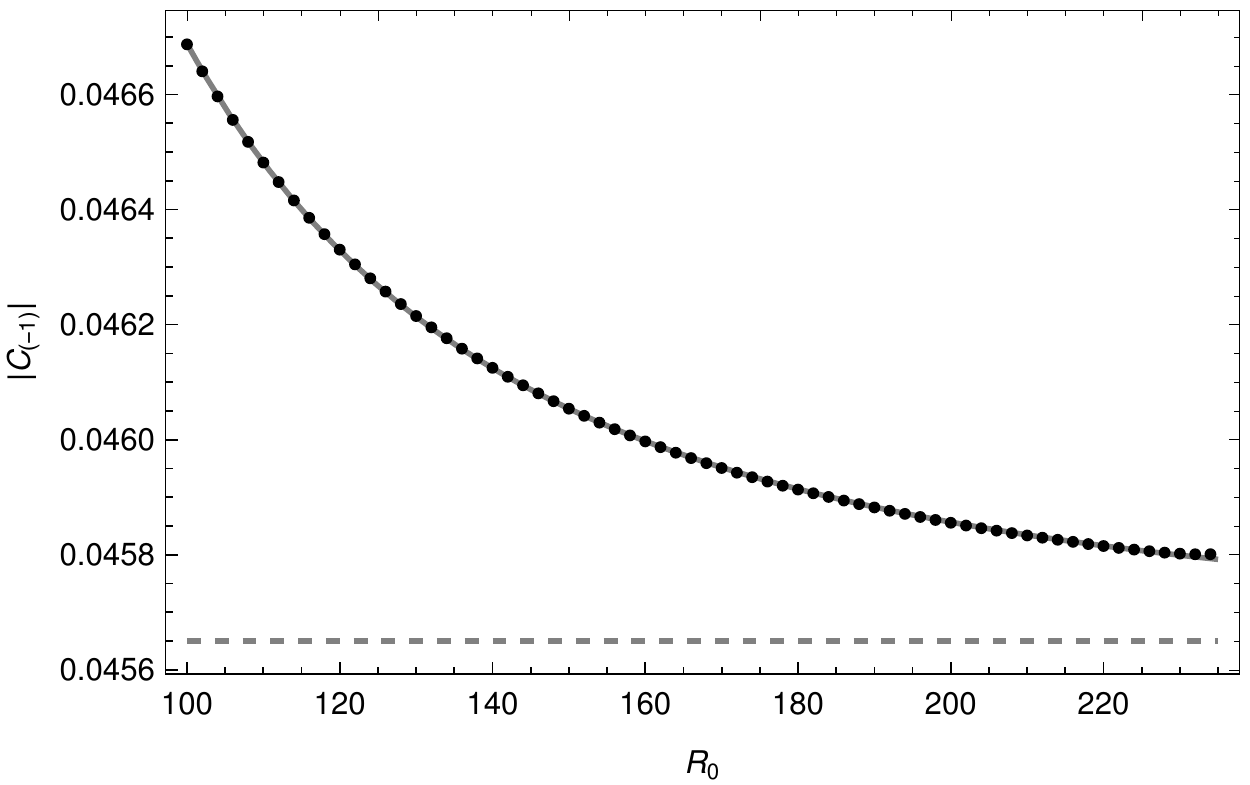}
	\centering
%	\captionsetup{width=.75\linewidth}
	\caption{\label{fig:M_asymptotic_C_-1}
		Points correspond to the coefficient $\vert c^{R_0}_{(-1)}\vert$ of the term proportional to $\frac{1}{m}$ in (\ref{fitmR0}). The solid line corresponds to the fit (\ref{eq:asymptotic_fit}) and the dotted one to the asymptotic value $|c_{(-1)}|=0.045659$ we take as the final result.}  
\end{figure}

Thus, the coefficients $c^{R_0}_{(-1)}$ obtained from (\ref{fitmR0}) through this procedure are still labeled with the corresponding $R_{0}$ identifying the lattice interval within which the calculation is done. 
The results are shown in figure \ref{fig:M_asymptotic_C_-1} where the points approach asymptotically a $c_{(-1)}$ value.  
To extract  the asymptotic value from the pairs $(R_0,c_{(-1)}^{R_0})$, we propose the fit
\begin{equation}\label{eq:asymptotic_fit}
	c_{(-1)}^{R_{0}}=c_{(-1)}+\alpha\, R_{0}^{-\beta}\,,
\end{equation}
which gives $c_{(-1)}=-0.045659$.

Finally, the Maxwell coefficient 
\begin{equation}\label{result_maxwell}
c^M_{log}=2c_{(-1)}\frac{L}{\pi R}=-0.0914\,\frac{L}{\pi R}\sim-\frac{7}{240}\frac{L}{R},\;
\end{equation}
with a relative error of $0.3\%$. Note that a factor two has been added in (\ref{result_maxwell}) due to the presence of two identical scalars in (\ref{eq:H_radial}).
\section{Conclusions}
Motivated by the mismatch between the EE logarithmic coefficient calculated in  \cite{dowker, coefsph} for the Maxwell field in the sphere and the expected one given by the type $a$ coefficient in the trace anomaly, we study here the EE for the Maxwell field in the cylinder. We choose this geometry to test the analytical prediction (\ref{coefcyl}) in \cite{solo}. The EE of  a cylindrical region is sensitive to the type $c$ anomaly for any conformal field theory in $(3+1)$ dimensions.
 The Maxwell Hamiltonian written in terms of the electric and magnetic physical fields $E$ and $B$, can be mapped to the Hamiltonian of two uncoupled scalar fields with a quadratic self interaction term (\ref{autointeraccion}). This mapping is possible using appropriate cylindrical coordinates and imposing periodic boundary conditions along the cylinder axis reducing the problem from the cylinder to the disk. The dimensional reduction, valid for free fields and cylinders with $L\gg R$, relates the logarithmic with the $\frac{1}{mR}$ coefficients of the cylinder and disk EE expansions respectively. In the dimensional reduced problem, the fields become massive with masses proportional to the momenta $k_n$ associated to each field mode $\phi_n(\rho,\theta)$.

Thus, we compute numerically the EE of a massive scalar field  plus the self interaction term in a disk. We obtain the Maxwell logarithmic coefficient for the cylinder is $c^M_{\log}\sim\frac{7}{240} L/R$ .  This disagrees with \cite{solo}, which set $c^M_{\log}=\frac{12}{240} L/R$ accordingly with the coefficient $c$ in the Maxwell trace anomaly. 
In the sphere case, studied in \cite{coefsph} it was also found a mismatch between the anomaly and the logarithmic coefficient. In this case, the entanglement entropy of a Maxwell field is equivalent to the one of two identical
massless scalars from which the  $l = 0$ mode has been removed. This shows the relation
$c_{\log}^M = 2(c_{\log}^S - c_{\log}^{S_{l=0 }})$ between the logarithmic coefficient in the entropy for a Maxwell field
$c_{\log}^M$ , the one for a $d = 4$ massless scalar $c_{\log}^S$, and the logarithmic coefficient $c_{\log}^{S_{l=0}}$ for a
$d = 2$ scalar with Dirichlet boundary condition at the origin. This gives $c_{\log}^M = −16/45$, which coincides with previous calculations \cite{dowker} and does not match the coefficient $−31/45$ in the  Maxwell field trace anomaly.
As we mention in the Introduction, the ambiguities in the assignation of local algebras to regions described in \cite{cashueros} and \cite{gaugecashue} could solve the EE logarithmic coefficient mismatch additioning a center on the boundary region. In fact, the classical boundary contribution that exactly gives the missing $-1/3$ in the spherical case corresponds to the electric center choice \cite{Maxwell,masmaxwell}. Moreover, in \cite{masmaxwell}, it is suggested that in general the classical contribution for any region is

\begin{equation}
S_{cl} = \frac{1}{2} \text{tr} \log (\nabla^2_{bd})\,,
\end{equation}
given the necessary non direct destilable contribution to the logarithmic EE coefficient to match the anomaly. However, also according to \cite{masmaxwell}, for a cylindrical region with zero intrinsic curvature this non destilable piece would give no contribution.

Nevertheless, as we understand the senseful well defined quantity for measuring entanglement for quantum fields is given by a \emph{regularized} EE defined in terms of the MI, and being this last not sensitive to the center content, the center choice cannot be the final solution to the mismatch problem.

\section*{Acknowledgments}
We thank H. Casini and G.Torroba for useful discussions and comments.
This work was partially supported by CONICET and Universidad Nacional de Cuyo, Argentina.

\bibliography{mibib}

\end{document}